\documentclass[a4paper,11pt]{article}
\usepackage{pos}
\usepackage{comment}
\usepackage{enumitem}

\usepackage{scalerel}
\usepackage{tikz}
\definecolor{orcidlogocol}{HTML}{A6CE39}
\title{Can KM3-230213A be compatible with a cosmogenic origin?}
\author*[a]{Antonio Condorelli}
\author[b,c]{Antonio Marinelli}
\onbehalf{on behalf of the KM3NeT collaboration}

\affiliation[a]{Universit{\'e} Paris Cit{\'e}, CNRS, Astroparticule et Cosmologie, F-75013 Paris, France}
\affiliation[b]{Universit{\`a} di Napoli ``Federico II'', Dip. Scienze Fisiche ``E. Pancini'', Complesso Universitario di Monte S. Angelo, Via Cintia ed. G, Napoli, 80126 Italy}
\affiliation[c]{INFN, Sezione di Napoli, Complesso Universitario di Monte S. Angelo, Via Cintia ed. G, Napoli, 80126 Italy}

\emailAdd{acondorelli@km3net.de}

\abstract{On the 13th February 2023 the KM3NeT/ARCA telescope observed a track-like event compatible with a ultra-high-energy muon with an estimated energy of 120 PeV, produced by a neutrino with an even higher energy, making it the most energetic neutrino event ever detected. The reported equivalent flux suggest the possible existence of a new diffuse component. A diffuse cosmogenic flux is expected to originate from the interactions of ultra-high-energy cosmic rays with ambient photon and matter fields. Here we show that this component can be compatible with the reported flux level only integrating the cosmogenic emission, at least up to redshift ~$z = 6 $  and assuming a subdominant fraction of protons in the ultra-high-energy cosmic-ray flux, thus placing constraints on known cosmic accelerators.
These conditions impose constraints on known cosmic accelerators and open a window into an unexplored region of the Universe at this energy scale.}

\ConferenceLogo{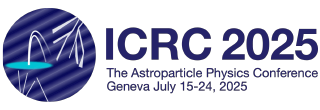}

\FullConference{39th International Cosmic Ray Conference (ICRC2025)\\
 15–24 July 2025\\
Geneva, Switzerland\\}
\begin{document}
\maketitle

\section{Introduction}
In recent decades, the discovery of high-energy cosmic messengers, i.e. gamma rays, cosmic rays, and neutrinos has confirmed the presence of extreme astrophysical accelerators in the Universe~\citep{IceCube:2013low,Fermi-LAT:2009ihh,PierreAuger:2015eyc}. To better understand these sources and the physical conditions surrounding them, a \textit{multi-messenger approach} has been adopted, combining detections from various messengers. However, the distant Universe remains largely inaccessible at the highest energies due to the attenuation of gamma rays and cosmic rays over cosmological distances.

Gamma rays above a few TeV interact with low-energy photons from the extragalactic background light (EBL)~\citep{Dominguez:2010bv} and the cosmic microwave background (CMB)~\citep{Planck:2018vyg}, resulting in $e^+e^-$ pair production, which significantly limits their travel range. Similarly, ultra-high-energy (UHE) cosmic rays ($\geq 10^{18} \ \rm eV$) undergo energy losses and photo-disintegration via interactions with CMB and EBL photons, limiting their mean free path to a few gigaparsecs (redshift $z \simeq 1$). At lower energies, magnetic field deflections further obscure their sources.

In contrast, neutrinos interact only weakly with matter and radiation, allowing them to propagate virtually unimpeded over cosmological distances. This makes them ideal probes of the most distant and obscured regions of the Universe. The detection of UHE neutrinos offers a unique confirmation of extreme cosmic accelerators located deep in the Universe.

This study, based on \citep{KM3NeT:2025vut}, investigates whether the neutrino event KM3-230213A, observed by the KM3NeT/ARCA detector with an estimated energy range of 120 PeV to 2.6 EeV (90\% CL), could be a \textit{cosmogenic neutrino}, produced by cosmic-ray interactions with background photons in the distant Universe.

\section{Cosmogenic neutrinos}
Ultra-high-energy cosmic rays (UHECRs) have been observed with energies exceeding \(10^{20}~\mathrm{eV}\), though their sources remain uncertain. These particles interact with cosmic background photons through processes such as pair production, photo-pion production (for protons), and photo-disintegration (for nuclei), leading to the creation of pions. The decay of charged pions and muons subsequently produces cosmogenic neutrinos.

These neutrinos primarily originate from interactions between cosmic rays and background photons across the Universe. The predicted neutrino flux strongly depends on the composition of cosmic rays, their maximum acceleration energy, the evolution of their sources, and the spectral shape. Early models with a pure-proton composition predicted fluxes on the order of \(E^2\phi_\nu \sim 10^{-9}~\mathrm{GeV~cm^{-2}~sr^{-1}~s^{-1}}\), but more recent mixed-composition models yield much lower predictions, around \(10^{-12}~\mathrm{GeV~cm^{-2}~sr^{-1}~s^{-1}}\), consistent with current observational constraints from the Pierre Auger Observatory.

Moreover, interactions of UHECRs with interstellar matter in the Galactic disk also produce cosmogenic neutrinos, providing a guaranteed minimal contribution to the overall flux.
\section{KM3$-$230213A}
\label{sec:Evt}

An ultra-high-energy muon event, designated KM3$-$230213A, was detected in the KM3NeT/ARCA detector on February 13, 2023, at 01:16:47 UTC. At the time of the event, 21 detection units were operational, and the detector in this configuration—referred to as ARCA21—had been collecting data since September 23, 2022. This configuration continued until September 11, 2023, yielding a total livetime $T^{\rm ARCA21}$ of 287.4 days. During this period, approximately 110 million events were triggered, among which KM3$-$230213A stands out as the highest-energy event observed.

The reconstructed muon energy for this event is estimated at $120^{+110}_{-60}$~PeV, with a 90\% confidence level interval spanning from 35 to 380~PeV. According to ARCA detector simulations, the median neutrino energy corresponding to such a muon is 220~PeV. Furthermore, 68\% (90\%) of the simulated neutrino events producing similar muons fall within the energy range of 110--790~PeV (72~PeV--2.6~EeV).

\section{Description of Cosmogenic Scenario}
\label{sec:framework}

\subsection{Injected Cosmic Ray Composition}

The acceleration of various particle species in astrophysical sources is typically modeled through non-thermal processes, characterized by power-law energy spectra. An exponential cutoff is introduced to describe the end of the acceleration mechanism, in the absence of strong theoretical constraints. For simplicity, the injected composition is grouped into five representative, stable nuclei: hydrogen ($^1$H), helium ($^4$He), nitrogen ($^{14}$N), silicon ($^{28}$Si), and iron ($^{56}$Fe), reflecting the expected diversity of accelerated nuclei.

The injection rate $q_{A}(E)$, defined as the number of nuclei with mass number $A$ emitted per unit comoving volume and per unit nucleon energy, is given by:
\begin{equation}
\label{eqn:qA}
    q_{A}(E) = q_{0A} \left( \frac{E}{E_0} \right)^{-\gamma_{A}} f(E, Z_{A}),
\end{equation}
where $\gamma_{A}$ is the spectral index and $q_{0A}$ denotes the normalization of the injection rate. The suppression function $f(E, Z)$, which models the high-energy cutoff:
\begin{equation}
\label{eqn:fsupp}
    f(E,Z) = 
    \begin{cases}
    1 & \text{if } E \leq E^{Z}_{\mathrm{max}}, \\
    \exp{\left(1 - E / E^{Z}_{\mathrm{max}} \right)} & \text{otherwise},
    \end{cases}
\end{equation}
where $E^{Z}_{\mathrm{max}} = Z E_{\mathrm{max}}$ expresses the scaling of the maximum acceleration energy with the particle’s electric charge $Z$, and $E_{\mathrm{max}}$ is a universal cutoff parameter.

For each nuclear species $A$, the differential energy production rate per unit comoving volume—directly linked to the source luminosity—is defined as $\ell_A(E,z) = E^2 q_A(E) S(z)$, where $S(z)$ represents the redshift evolution of the UHECR luminosity density. The bolometric energy production rate per unit comoving volume at redshift $z$ is then computed as:
\begin{equation}
    \mathcal{L}_A(E, z) = S(z) \int_{E}^{\infty} \mathrm{d}E' \, E' \, q_A(E').
    \label{emissivity}
\end{equation}

\subsection{Cosmological Source Evolution}

The ultra-high-energy cosmic rays detected by the Pierre Auger Observatory \citep{PierreAuger:2021hun} and the Telescope Array \citep{ABBASI2023102864}, along with their observed spectra, are predominantly influenced by nearby astrophysical accelerators. Consequently, the properties of UHECR sources at higher redshifts ($z \gtrsim 1$) remain largely unconstrained. Various classes of astrophysical objects—both steady and transient—are potential candidates for accelerating cosmic rays up to energies of $10^{21}$ eV, particularly within extreme environments. \\
The redshift evolution of the comoving source density is commonly parametrized as:
\begin{equation}
S(z) \propto (1 + z)^{m},
\label{eqn:cosmditr}
\end{equation}
where the exponent $m$ governs the strength and direction of the evolutionary trend. In this work, $m$ is varied from $-5$ to $5$ in steps of 0.2, encompassing a wide range of possible evolution scenarios without committing to any specific source class.

The resulting flux of cosmogenic neutrinos is sensitive to source distance and, therefore, to the assumed source evolution model. The emission rate density $L(E,z)$ of cosmic rays per unit comoving volume for a given source population is modeled as:
\begin{equation}
L(E,z) = S(z) \times Q_{\rm CR}(E),
\label{eqn:Lumin}
\end{equation}
where $Q_{\rm CR}(E)$ denotes the cosmic-ray injection term specific to the source class, and is defined as the sum over the injection rates $q_A(E)$ for each nuclear species introduced in Equation~\ref{eqn:qA}.

\section{Expected Neutrino Fluxes for Different Redshifts}
\label{sec:results}

\begin{figure}[t]
    \centering
    \includegraphics[width=0.8\linewidth]{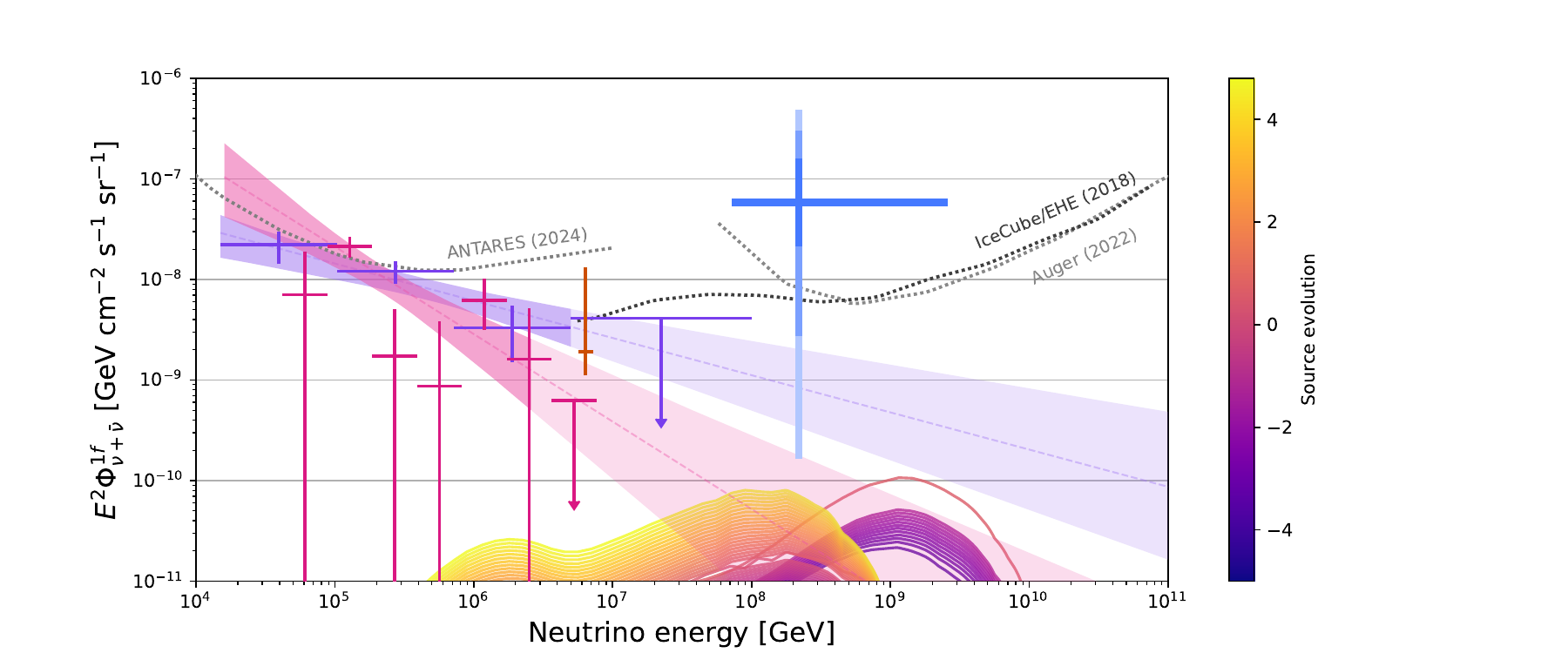} 
    \includegraphics[width=0.8\linewidth]{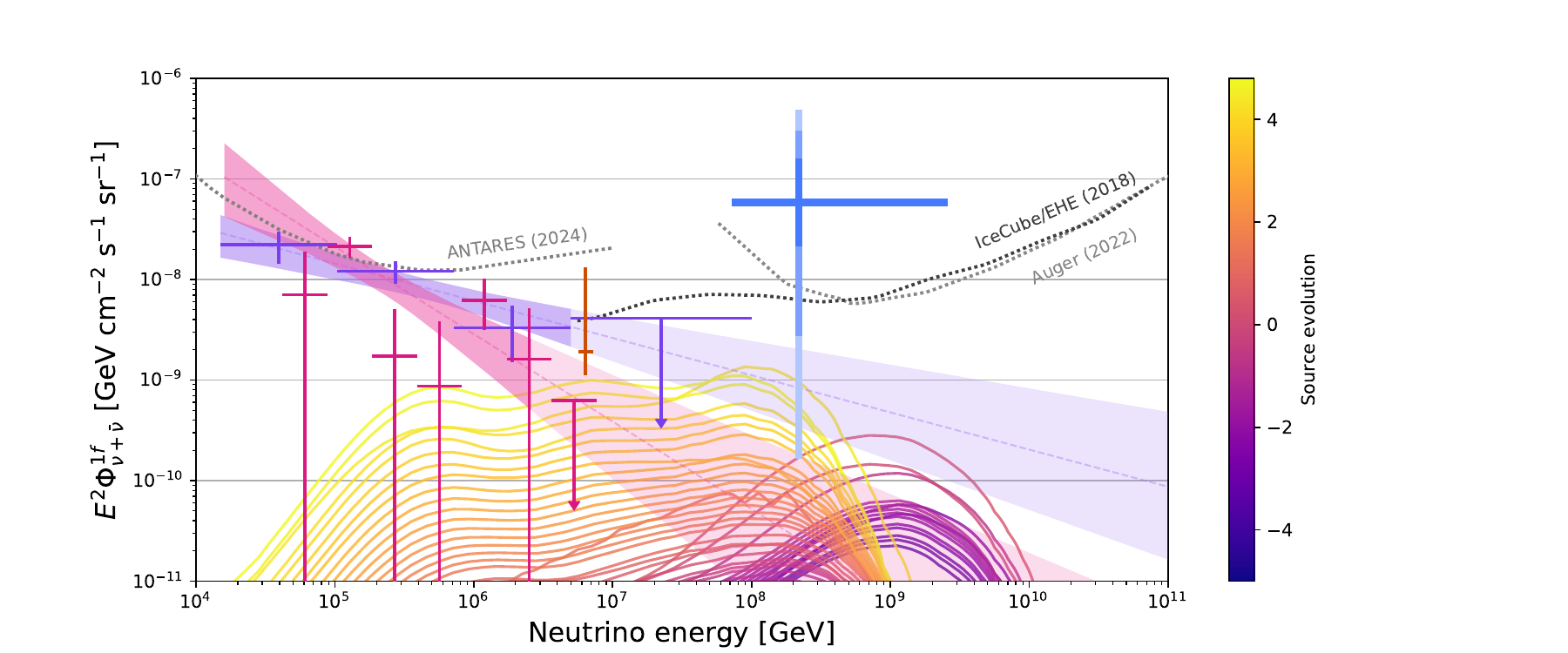} 
    \caption{Expected neutrino fluxes as a function of energy for different source evolution parameters ($m$; color-coded) at two maximum redshift values: $z_{\rm max} = 1$ (top panel) and $z_{\rm max} = 6$ (bottom panel).\\
    The blue cross indicates the flux required to produce one expected event within the central 90\% CL energy range of the KM3-230213A event (horizontal span); vertical bars denote the 1$\sigma$, 2$\sigma$, and 3$\sigma$ Feldman-Cousins confidence intervals \citep{KM3NeT2024}. The purple- and pink-shaded bands represent 68\% confidence level contours of IceCube's single power-law fits (Northern-Sky Tracks and High-Energy Starting Events, respectively). Darker regions indicate the 90\% central energy range at the best fit (dashed lines), while lighter shades are extrapolations to higher energies. Purple and pink crosses represent the respective IceCube best fits; the orange cross corresponds to the Glashow resonance event. Dotted lines show upper limits from ANTARES (95\% CL), Pierre Auger, and IceCube.}
    \label{fig:candidate}
\end{figure}
 The resulting cosmogenic neutrino fluxes, obtained using the best-fit parameters that describe the observed UHECR spectrum and composition, are presented in Figure~\ref{fig:candidate}.

In the top panel, the cosmic-ray sources are restricted to a maximum redshift $z_{\rm max} = 1$, consistent with the expectation that UHECRs above the ankle experience significant energy loss over large distances. The bottom panel extends the distribution to $z_{\rm max} = 6$.

In both cases, the neutrino fluxes exhibit two distinct peaks: one at lower energies due to interactions with the extragalactic background light (EBL), and one at higher energies resulting from interactions with the CMB. A clear trend is observed when transitioning from negative to positive source evolution (depicted from violet to yellow tones). This variation arises due to changes in the UHECR composition and propagation:

\begin{itemize}
    \item \textbf{Negative evolution ($m < 0$):} Nearby sources dominate, reducing opportunities for photodisintegration of heavy nuclei. Consequently, the injected composition resembles the one observed at Earth, favoring intermediate-mass nuclei and resulting in low neutrino fluxes.
    \item \textbf{Positive evolution ($m > 0$):} Distant sources dominate, allowing extensive propagation and disintegration. Heavier nuclei at injection can transform into lighter species and protons, which are crucial for neutrino production. As a result, cosmogenic neutrino fluxes are significantly enhanced in these scenarios.
\end{itemize}

In both redshift cases ($z_{\rm max} = 1$ and $z_{\rm max} = 6$), positive source evolution produces higher neutrino fluxes due to the increased contribution from secondary protons, which play a pivotal role in neutrino production.\\
Notably, for $z_{\rm max} = 1$, fluxes remain below $10^{-10} \, \mathrm{GeV \, cm^{-2} \, s^{-1} \, sr^{-1}}$ in the relevant energy range. In contrast, extending to $z_{\rm max} = 6$ increases the flux by up to an order of magnitude. This suggests that, if the KM3-230213A event is of cosmogenic origin, it is more likely associated with distant accelerators rather than nearby UHECR sources.\\
To quantify this, the expected number of events $n_{\rm exp}$ in ARCA21 for a given flux model $\Phi_{\rm model}$ is computed as:

\begin{equation}
    \begin{split}
        n_{\rm exp}^{\rm ARCA21} &= 
        T^{\rm ARCA21} \sum_i 
        \int_{\Delta\Omega_i} \int_{\Delta E} 
        A_{\rm eff}^{\rm ARCA21}(E, \Omega) \\
        &\quad \times \Phi_{\rm model}(E, \Omega) \, 
        {\rm d}E \, {\rm d}\Omega,
    \end{split}
    \label{eq:eff_modified}
\end{equation}

where $A_{\rm eff}^{\rm ARCA21}(E, \Omega)$ is the effective area for a given energy and zenith angle bin $\Delta\Omega_i$, and $T^{\rm ARCA21}$ is the live time. The calculation follows the same methodology as in \citep{KM3NeT2024}, integrating above $100 \, \mathrm{PeV}$ using the same event selection criteria.

\begin{figure}[t]
    \centering
    \includegraphics[width=0.7\linewidth]{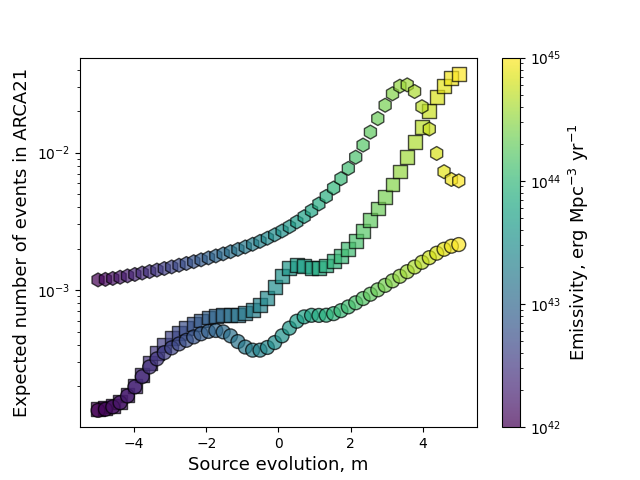} 
    \caption{Expected number of events in ARCA21 as a function of source evolution $m$ for $z_{\rm max} = 1$ (circles) and $z_{\rm max} = 6$ (squares). The color scale represents the corresponding cosmic-ray emissivity required to match the UHECR spectrum at $z = 0$, as defined in Equation~\ref{emissivity}.}
    \label{fig:evol}
\end{figure}

Using Equation~\ref{eq:eff_modified}, we estimate the expected number of events in ARCA21 across the scanned values of $m$ for both $z_{\rm max} = 1$ and $z_{\rm max} = 6$. Results are shown in Figure~\ref{fig:evol}. For negative $m$, the expected number of events remains low and nearly identical in both redshift scenarios. As $m$ increases, the divergence between the two redshift cases becomes more pronounced, with higher $z_{\rm max}$ producing more neutrinos.

The color axis of Figure~\ref{fig:evol} shows the cosmic-ray emissivity, defined as the luminosity density required to sustain the UHECR spectrum at $z = 0$ (see Equation~\ref{emissivity}). As expected, models with positive $m$ values demand higher emissivity due to increased interaction rates and neutrino production.

Identifying plausible UHECR accelerators based on known extragalactic objects requires consideration of their physical properties inferred from electromagnetic observations. While these properties may be approximately constant over cosmic time, probing their acceleration efficiency and redshift evolution directly through high-energy messengers remains essential. The KM3-230213A observation enables exploration of a previously untested region of the Universe in the $10^{17}$–$10^{18}$ eV neutrino energy range, corresponding to UHECRs of up to $10^{20} \, \mathrm{eV}$—an energy domain beyond the reach of Pierre Auger and Telescope Array.

It would be informative to compare the emissivity values shown in Figure~\ref{fig:evol} with those of candidate source populations. However, the plotted emissivities represent the energy injection in \textit{cosmic rays}, whereas source emissivity estimates from the literature are typically in terms of electromagnetic or gamma-ray output. A direct comparison thus requires assumptions about the cosmic-ray loading factor, i.e., the fraction of energy allocated to hadronic cosmic rays relative to electromagnetic components.
\section{Constraining the proton fraction at the highest energies}
\label{sec:6}

Protons are considerably more efficient than nuclei at producing pions, and consequently high-energy neutrinos—due to their effective interactions with photon fields, primarily the CMB at the highest energies, via photo-pion production (\(p + \gamma \rightarrow \Delta^+ \rightarrow \pi + N\)). The charged pions produced in this process subsequently decay into neutrinos.  
Although photo-pion production also occurs for nucleons bound within ultra-high-energy (UHE) nuclei, where the interacting nucleon is ejected from the parent nucleus, this process is generally subdominant compared to nucleus photo-disintegration except at extremely high energies.

Recent studies \citep{Decerprit:2011qe, Muzio:2019leu, PierreAuger:2022atd} have demonstrated that including a proton component, while preserving the expected mass composition dominated by intermediate masses, can significantly increase the predicted neutrino flux.\\

Within the cosmogenic framework, we explore this scenario by introducing a secondary, proton-only component characterized by a distinct spectral index. This component is fitted to reproduce the proton fraction measured by the Pierre Auger Observatory by matching the energy spectrum and mass composition data below the ankle and then extrapolating to the highest energies.

\begin{figure}[h!]
    \centering
    \includegraphics[width=0.8\textwidth]{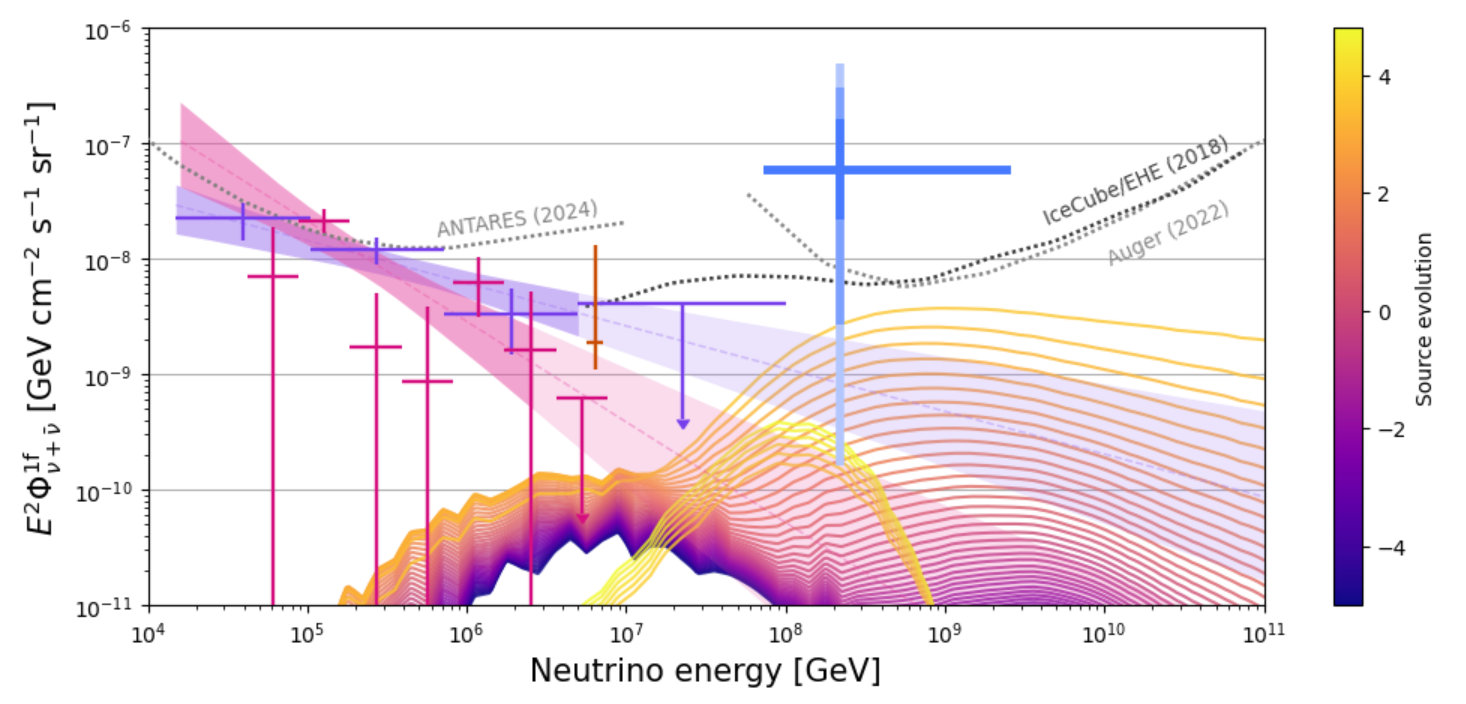} 
    \caption{Expected neutrino fluxes as a function of energy assuming a secondary proton component subdominant at the highest energies. The format follows that of figure~\ref{fig:candidate}.}
    \label{fig:proton}
\end{figure}

Figure~\ref{fig:proton} presents the resulting neutrino flux predictions, illustrating how the presence of even a subdominant proton fraction can enhance the expected neutrino flux.

In conclusion, even a small proton fraction at the highest energies leads to a substantial increase in the cosmogenic neutrino flux, improving prospects for future detections.  
The observation of KM3-230213A supports the existence of such a subdominant proton component at the highest energies, assuming a cosmogenic origin.

\newpage
\tikzset{
orcidlogo/.pic={
\fill[orcidlogocol] circle [radius=0.18];
\node[white,font=\sffamily\bfseries\scriptsize] at (0,0) {iD};
}
}

\newcommand{\orcidlink}[1]{%
\href{https://orcid.org/#1}{%
\scalerel*{%
\tikz[baseline=-0.1ex]{\pic{orcidlogo}}%
}{X}%
}%
}

\section*{The KM3NeT Collaboration}
\scriptsize
\begin{sloppypar}\noindent

%% 
%% Copyright 2007, 2008, 2009 Elsevier Ltd
%% 
%% This file is part of the 'Elsarticle Bundle'.
%% ---------------------------------------------results are shown
%% 
%% It may be distributed under the conditions of the LaTeX Project Public
%% License, either version 1.2 of this license or (at your option) any
%% later version.  The latest version of this license is in
%%    http://www.latex-project.org/lppl.txt
%% and version 1.2 or later is part of all distributions of LaTeX
%% version 1999/12/01 or later.
%% 
%% The list of all files belonging to the 'Elsarticle Bundle' is
%% given in the file `manifest.txt'.
%% 
%% Template article for Elsevier's document class `elsarticle'
%% with numbered style bibliographic references
%% SP 2008/03/01
% ----- Start automatically generated KM3NeT info
% ----- Start author list
%\cortext[cor]{corresponding author}
{O.~Adriani\,\orcidlink{0000-0002-3592-0654}}$^{b,a}$,
{S.~Aiello}$^{c}$,
{A.~Albert}$^{d,bd}$,
{A.\,R.~Alhebsi\,\orcidlink{0009-0002-7320-7638}}$^{e}$,
{M.~Alshamsi}$^{f}$,
{S. Alves Garre\,\orcidlink{0000-0003-1893-0858}}$^{g}$,
{A. Ambrosone}$^{i,h}$,
{F.~Ameli}$^{j}$,
{M.~Andre}$^{k}$,
{L.~Aphecetche\,\orcidlink{0000-0001-7662-3878}}$^{l}$
% ...
\\ \\
% ----- End author list
% ----- Start address list
$^{a}${Universit{\`a} di Firenze, Dipartimento di Fisica e Astronomia, via Sansone 1, Sesto Fiorentino, 50019 Italy}\\
$^{b}${INFN, Sezione di Firenze, via Sansone 1, Sesto Fiorentino, 50019 Italy}\\
$^{c}${INFN, Sezione di Catania, (INFN-CT) Via Santa Sofia 64, Catania, 95123 Italy}\\
$^{d}${Universit{\'e}~de~Strasbourg,~CNRS,~IPHC~UMR~7178,~F-67000~Strasbourg,~France}\\
$^{e}${Khalifa University of Science and Technology, Department of Physics, PO Box 127788, Abu Dhabi,   United Arab Emirates}\\
$^{f}${Aix~Marseille~Univ,~CNRS/IN2P3,~CPPM,~Marseille,~France}\\
$^{g}${IFIC - Instituto de F{\'\i}sica Corpuscular (CSIC - Universitat de Val{\`e}ncia), c/Catedr{\'a}tico Jos{\'e} Beltr{\'a}n, 2, 46980 Paterna, Valencia, Spain}\\
$^{h}${INFN, Sezione di Napoli, Complesso Universitario di Monte S. Angelo, Via Cintia ed. G, Napoli, 80126 Italy}\\
$^{i}${Universit{\`a} di Napoli ``Federico II'', Dip. Scienze Fisiche ``E. Pancini'', Complesso Universitario di Monte S. Angelo, Via Cintia ed. G, Napoli, 80126 Italy}\\
$^{j}${INFN, Sezione di Roma, Piazzale Aldo Moro, 2 - c/o Dipartimento di Fisica, Edificio, G.Marconi, Roma, 00185 Italy}\\
$^{k}${Universitat Polit{\`e}cnica de Catalunya, Laboratori d'Aplicacions Bioac{\'u}stiques, Centre Tecnol{\`o}gic de Vilanova i la Geltr{\'u}, Avda. Rambla Exposici{\'o}, s/n, Vilanova i la Geltr{\'u}, 08800 Spain}\\
$^{l}${Subatech, IMT Atlantique, IN2P3-CNRS, Nantes Universit{\'e}, 4 rue Alfred Kastler - La Chantrerie, Nantes, BP 20722 44307 France}\\
% ...
$^{bd}${Universit{\'e} de Haute Alsace, rue des Fr{\`e}res Lumi{\`e}re, 68093 Mulhouse Cedex, France}\\
% ...
% ----- End address list
% ----- End automatically generated KM3NeT info
\end{sloppypar}

\section*{Acknowledgements}
The authors acknowledge the financial support of:
%INFRADEV
KM3NeT-INFRADEV2 project, funded by the European Union Horizon Europe Research and Innovation Programme under grant agreement No 101079679;
%Belgium
Funds for Scientific Research (FRS-FNRS), Francqui foundation, BAEF foundation.
%Czeck
Czech Science Foundation (GAČR 24-12702S);
%France
Agence Nationale de la Recherche (contract ANR-15-CE31-0020), Centre National de la Recherche Scientifique (CNRS), Commission Europ\'eenne (FEDER fund and Marie Curie Program), LabEx UnivEarthS (ANR-10-LABX-0023 and ANR-18-IDEX-0001), Paris \^Ile-de-France Region, Normandy Region (Alpha, Blue-waves and Neptune), France,
%For the CPER
The Provence-Alpes-Côte d'Azur Delegation for Research and Innovation (DRARI), the Provence-Alpes-Côte d'Azur region, the Bouches-du-Rhône Departmental Council, the Metropolis of Aix-Marseille Provence and the City of Marseille through the CPER 2021-2027 NEUMED project,
%For IN2P3
The CNRS Institut National de Physique Nucléaire et de Physique des Particules (IN2P3);
%Georgia
Shota Rustaveli National Science Foundation of Georgia (SRNSFG, FR-22-13708), Georgia;
%Germany (Max Planck Inst.)
This work is part of the MuSES project which has received funding from the European Research Council (ERC) under the European Union’s Horizon 2020 Research and Innovation Programme (grant agreement No 101142396).
%Greece
The General Secretariat of Research and Innovation (GSRI), Greece;
%Italy
Istituto Nazionale di Fisica Nucleare (INFN) and Ministero dell’Universit{\`a} e della Ricerca (MUR), through PRIN 2022 program (Grant PANTHEON 2022E2J4RK, Next Generation EU) and PON R\&I program (Avviso n. 424 del 28 febbraio 2018, Progetto PACK-PIR01 00021), Italy; IDMAR project Po-Fesr Sicilian Region az. 1.5.1; A. De Benedittis, W. Idrissi Ibnsalih, M. Bendahman, A. Nayerhoda, G. Papalashvili, I. C. Rea, A. Simonelli have been supported by the Italian Ministero dell'Universit{\`a} e della Ricerca (MUR), Progetto CIR01 00021 (Avviso n. 2595 del 24 dicembre 2019); KM3NeT4RR MUR Project National Recovery and Resilience Plan (NRRP), Mission 4 Component 2 Investment 3.1, Funded by the European Union – NextGenerationEU,CUP I57G21000040001, Concession Decree MUR No. n. Prot. 123 del 21/06/2022;
%Morocco
Ministry of Higher Education, Scientific Research and Innovation, Morocco, and the Arab Fund for Economic and Social Development, Kuwait;
%The Netherlands
Nederlandse organisatie voor Wetenschappelijk Onderzoek (NWO), the Netherlands;
%Poland
The grant “AstroCeNT: Particle Astrophysics Science and Technology Centre”, carried out within the International Research Agendas programme of the Foundation for Polish Science financed by the European Union under the European Regional Development Fund; The program: “Excellence initiative-research university” for the AGH University in Krakow; The ARTIQ project: UMO-2021/01/2/ST6/00004 and ARTIQ/0004/2021;
%Romania
Ministry of Research, Innovation and Digitalisation, Romania;
%Slovak Republic
Slovak Research and Development Agency under Contract No. APVV-22-0413; Ministry of Education, Research, Development and Youth of the Slovak Republic;
%Spain
MCIN for PID2021-124591NB-C41, -C42, -C43 and PDC2023-145913-I00 funded by MCIN/AEI/10.13039/501100011033 and by “ERDF A way of making Europe”, for ASFAE/2022/014 and ASFAE/2022 /023 with funding from the EU NextGenerationEU (PRTR-C17.I01) and Generalitat Valenciana, for Grant AST22\_6.2 with funding from Consejer\'{\i}a de Universidad, Investigaci\'on e Innovaci\'on and Gobierno de Espa\~na and European Union - NextGenerationEU, for CSIC-INFRA23013 and for CNS2023-144099, Generalitat Valenciana for CIDEGENT/2018/034, /2019/043, /2020/049, /2021/23, for CIDEIG/2023/20, for CIPROM/2023/51 and for GRISOLIAP/2021/192 and EU for MSC/101025085, Spain;
%UAE
Khalifa University internal grants (ESIG-2023-008, RIG-2023-070 and RIG-2024-047), United Arab Emirates;
%UK
The European Union's Horizon 2020 Research and Innovation Programme (ChETEC-INFRA - Project no. 101008324).

% Lista delle label usate:

\end{document}